# MAST Upgrade – Construction Status

Joe Milnes[a], Nizar Ben Ayed, Fahim Dhalla, Geoff Fishpool, John Hill, Ioannis Katramados, Richard Martin, Graham Naylor, Tom O'Gorman, Rory Scannell, and the MAST Upgrade team

[a]*Culham Centre for Fusion Energy, Culham, Oxfordshire, UK*

The Mega Amp Spherical Tokamak (MAST) is the centre piece of the UK fusion research programme. In 2010, a MAST Upgrade programme was initiated with three primary objectives, to contribute to: 1) Testing reactor concepts (in particular exhaust solutions via a flexible divertor allowing Super-X and other extended leg configurations); 2) Adding to the knowledge base for ITER (by addressing important plasma physics questions and developing predictive models to help optimise ITER performance of ITER) and 3) Exploring the feasibility of using a spherical tokamak as the basis for a fusion Component Test Facility.

With the project mid-way through its construction phase, progress will be reported on a number of the critical subsystems. This will include manufacture and assembly of the coils, armour and support structures that make up the new divertors, construction of the new set coils that make up the centre column, installation of the new power supplies for powering the divertor coils and enhanced TF coil set, progress in delivering the upgraded diagnostic capability, the modification and upgrading of the NBI heating systems and the complete overhaul of the machine control infrastructure, including a new control room with full remote participation facilities.

## 1. Introduction

As reported in [1], the first phase of an upgrade programme to the MAST tokamak is underway. An annotated cross section of the new machine is shown in figure 1. This first phase will deliver:

1. A new, highly flexible divertor able to investigate long-legged configurations including Super-X [2,3], snowflake [4] and others.

2. Longer pulse capability (0.5 to 5s)

3. Increased Toroidal Field (0.5 to 0.8T at R = 0.8m)

4. Increased Plasma Current (1 to 2MA)

5. Off-axis Neutral Beam Injection

6. Advanced divertor diagnostic capabilities

The design phase of the key mechanical systems is largely complete and the project is now mid-way through the strip out / rebuild phase. A summary of the project timeline is shown in Table 1.

Table 1. Project timeline summary

| Milestone | Date |
|---|---|
| Conceptual Design started[a] | January 2008 |
| Funding approved | April 2010 |
| Strip out of MAST started | October 2013 |
| Rebuild of MAST-U started | March 2014 |
| Pump down (expected) | June 2015 |
| First Plasma (expected) | January 2016 |

*a: MAST-U concept with a Super-X divertor*

Once complete, this machine will become a key part of the EU's Medium Sized Tokamak (MST) programme, along with TCV and ASDEX Upgrade. Key features that will contribute to MAST-U's unique capability include a plasma facing geometry that enables a wide range of strike point locations, divertor chamber segregation enabling closure at the X-point, a variable speed cryopump, tight control of coil alignment and magnetic imperfections to minimize error fields, a large suite of state-of-the-art diagnostics and a highly flexible power supply network.

These and other features will be illustrated in the following section where key areas of progress are reported, including some critical design challenges that have been overcome and lessons learned along the way.

## 2. Construction Status

### 2.1 General

Construction of the original MAST device was completed in 2000 and its final operating campaign was completed in September 2013. At that time, the facility was handed over to the upgrade team to enable the strip out activities to begin.

Using a pre-constructed 50 Tonne crane (see figure 2) and detailed planning, the strip out and transfer of the MAST vacuum vessel to its assembly area was completed in less than 3 months, 1 week ahead of schedule. During this phase, a subset of the legacy systems, including many of the diagnostics and heating systems, were removed to their own dedicated assembly areas, inspected, refurbished and stored, ready for the rebuild phase expected to be completed in mid-2015.

_________________________________________

*author's email: joe.milnes@ccfe.ac.uk*

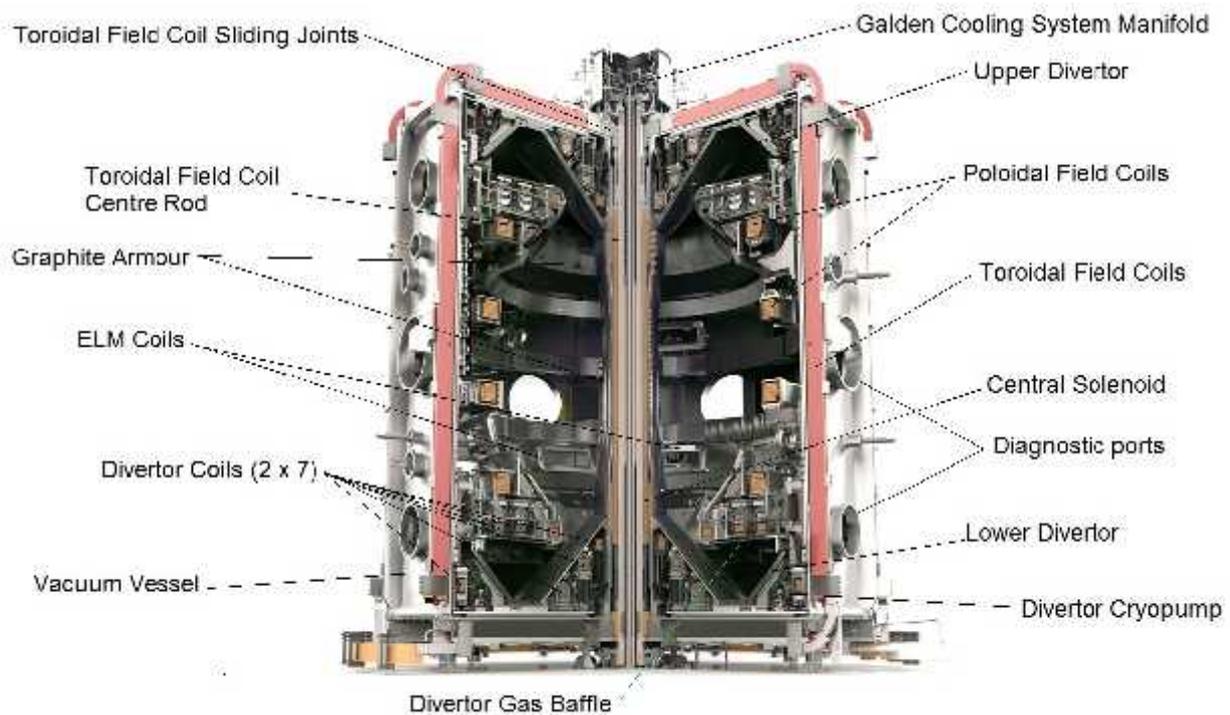

Fig. 1. Annotated Cross-Section of the MAST-U machine

In early 2014, there followed a 3 month window to complete the modifications to the vacuum vessel and preparation of many of the assembly and operational areas required for the new machine. In particular, a total of 50 new penetrations were precisely machined into the vacuum vessel to allow new ports to be welded into position. These will support the large number of additional diagnostics as well as provide feedthroughs for the 32 in-vessel coils (including ELM coils), gas introduction system and pumping systems required by the new machine.

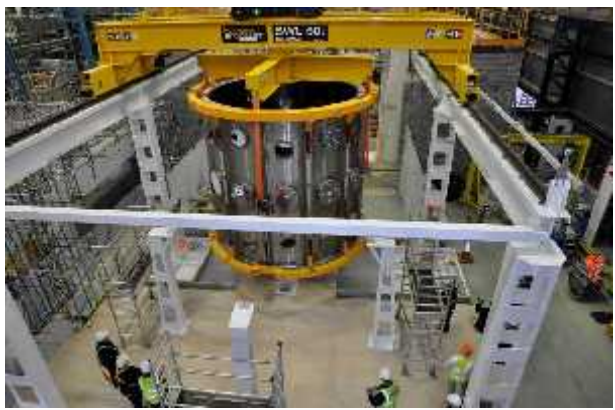

Fig. 2. MAST Vacuum Vessel being craned into its assembly area.

### 2.2 Load Assembly

The largest and most complex part of the Upgrade is the MAST-U Load Assembly (which refers to all the systems up to and including the MAST-U primary vacuum vessel).

Over 100,000 new components are required to complete the Load Assembly with only the original vacuum vessel itself and 2 pairs of the original MAST coils being re-used. At the time of writing, the following manufacturing progress can be reported:

- 12 of the 16 new in-vessel coils have been delivered.

- 3 out of the 4 new ex-vessel coils, including the new, enhanced solenoid (see figure 3), have been wound and impregnated

- 60% of the 388 divertor armour tiles have been delivered (there are also a further ~1,000 protection tiles, all of which are in manufacture).

- The High Field Side Gas Introduction System and both cryopumps have been fully assembled and are available for installation.

- In all, over 80,000 individual components have been delivered to site

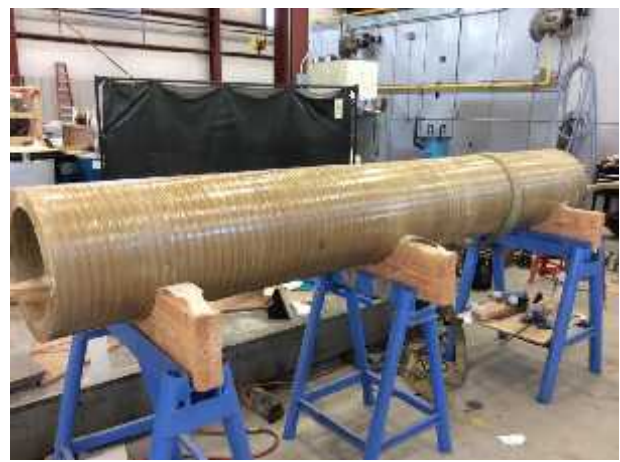

Fig. 3. New Solenoid

To meet the challenging timescales the project has been given to build this new machine, the project team has designed the Load Assembly be broken down and assembled into discrete modules. The primary modules are the lower and upper end plate modules, the outer cylinder module, the centre tube and centre column modules. These are illustrated in figure 4. A recent decision to split the outer cylinder into 3 futher modules (outer cylinder + lower and upper nose cassettes) takes the total number of Load Assembly work faces to 7 (the only way this machine could possibly be built in the targeted timeframe)

in preparation for the divertor armour and carriers (see figure 7)

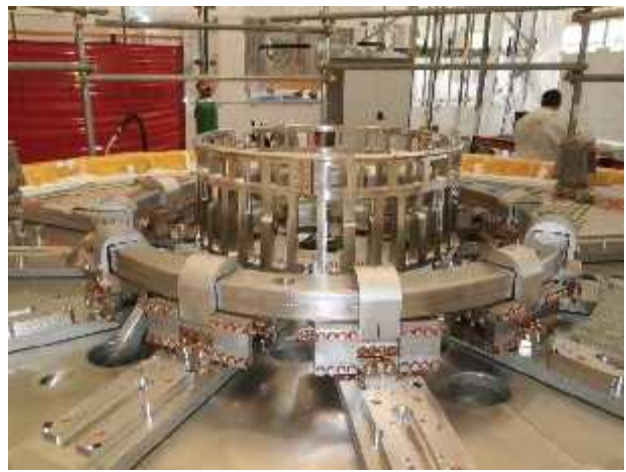

Fig. 5. Lower End Plate module, with D3 coil and major support ring installed

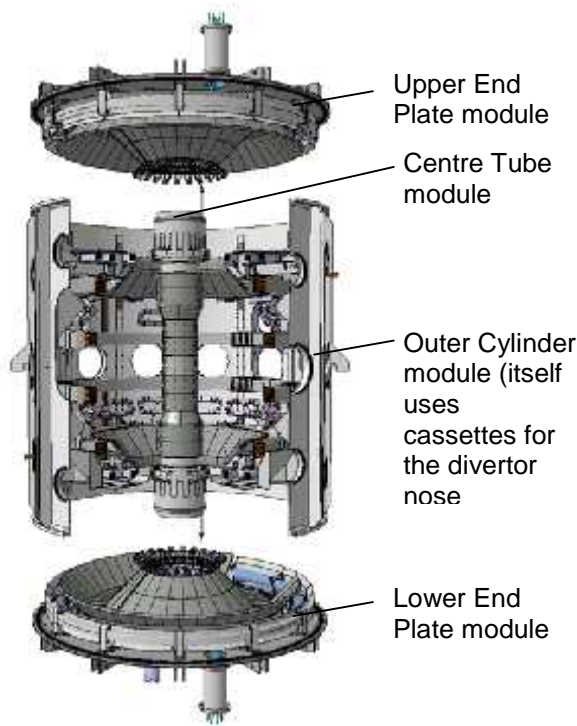

Fig. 4. Load Assembly Modular Design

Over the past 12 months, unforeseen design complexity has meant the design of the MAST-U Load Assembly has been delayed somewhat (some of the design challenges are summarised in the paragraphs below). This has resulted in many of the modules now working to a "Just In Time" delivery philosophy with respect to the new hardware being procured. This requires careful planning to minimize slippage to the overall programme when inevitable delivery delays occur.

Despite this, the build schedule currently shows limited slippage compared to the original target pump down date of April 2015 and a number of complex components have already been built, tested and installed into their respective modules. These include the following:

• Numerous coils and their respective support structures installed on both the End Plate and Outer Cylinder cassette modules (see figures 5 and 6)

• The major divertor amour support rings installed on the Lower End Plate divertor (see figure 5),

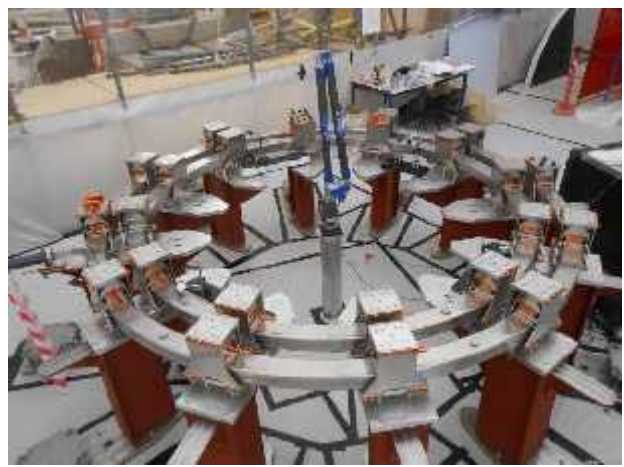

Fig. 6. Outer Cylinder Divertor Nose Cassette build status

The progress made to date has only been accomplished via a concerted effort from the design, manufacturing and installation teams to solve some critical engineering challenges. These include late modification of the support structure design to include mechanisms to facilitate maintenance from the plasma facing side (see figure 7), the design and assembly of coil supports required to sustain loads of many tonnes in one direction without over-constraining the coil in the radial direction (to allow for bakeout, see figure 8), the 3D profiling of the divertor plasma facing surfaces to maximize performance (developing methodologies to link physics codes to CAD output [5]), developing specific heat treatment methodologies and selection of manufacturing methods to minimize magnetic permeability [6] and designing a gas baffle between the primary chamber and divertor chambers with sufficiently low conductance.

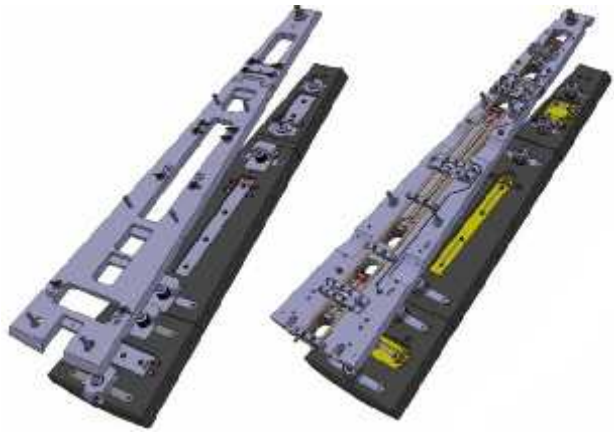

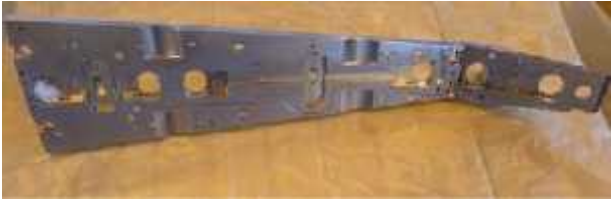

Fig. 7. Divertor Armour Slat Support (top left = original, top right = final, bottom = prototype)

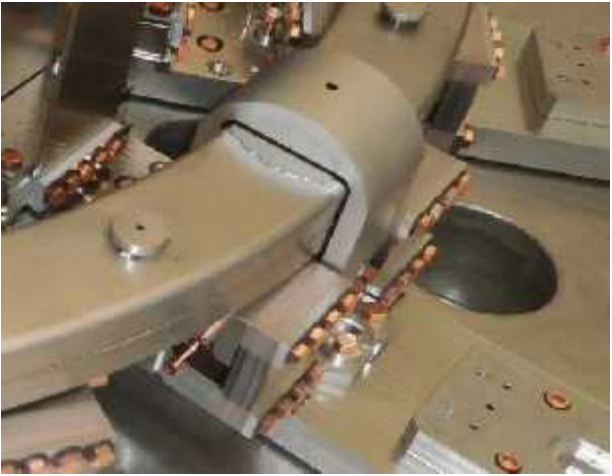

Fig. 8. Divertor Coil Support (use of lamellae for orthotropic stiffness)

### 2.3 Power Supplies

The MAST-U Power Supplies make up 20% of the total £34M budget (and over 30% of total hardware budget). This investment will significantly improve both the performance and reliability of the new machine.

In 2012, a concerted effort was made by the design team and the MAST-U physicists to define specifications for the new TF and Divertor Coils power supplies despite the full operational envelope of the machine still under review. As a result, these were some of the first contracts placed by the project team and by working closely with the suppliers to resolve a number of key technical issues, these two systems were delivered within a month or so of the original target (see figure 9).

At the time of writing, ongoing power supply procurements and installation work was broadly on track and it is expected that dummy load commissioning will be completed in time for integrated "power-supplies-into-coil" commissioning expected mid 2015.

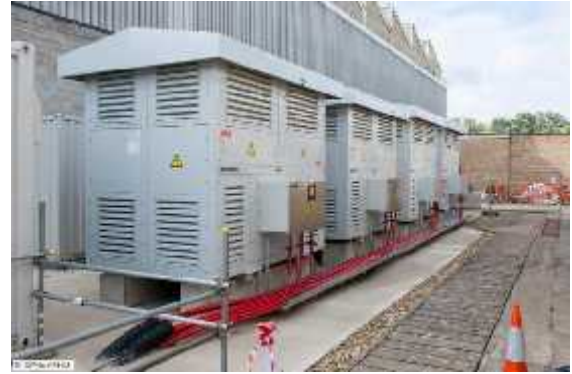

Fig. 9. New Toroidal Field Power Supplies Rectifier Cubicles (x4)

### 2.4 Diagnostics

Like its predecessor, MAST-U will have a significant suite of diagnostics with over 50 discrete systems envisaged. These will be made up of original MAST diagnostics (including reciprocating probes, IR cameras and Divertor Science Facility) as well as a number of new systems. These include over 600 Langmuir Probes, 500 Pick Up coils, a new $CO_2$ Interferometer (see figure 10), new bolometers (in both the main and divertor chambers) and a divertor Thomson Scattering system (see figure 11)

The manufacturing, assembly and testing of most of these systems is on schedule to ensure the early experimental programme envisaged in early 2016 is as productive as possible.

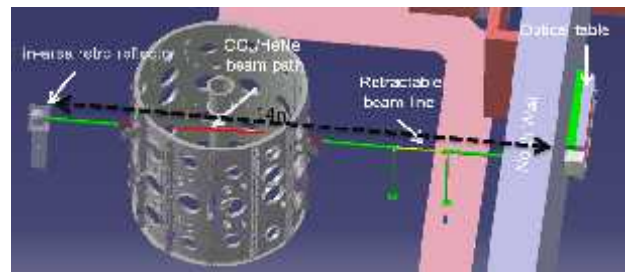

Fig. 10. New MAST-U $CO_2$ Interferometer

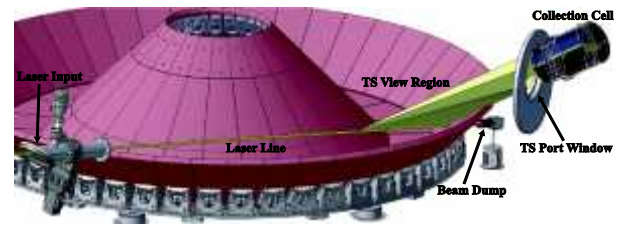

Fig. 11. New MAST-U Divertor Thomson Scattering system

### 2.5 Heating and Fuelling

As for its predecessor, the auxiliary heating for MAST-U will be provided by Neutral Beam Injection. As stated in [7], "a project re-scoping exercise in 2011 (triggered by a detailed budget review) postponed the installation of the new Double Beam Box (DBB) described in [8] that would have increased the total NBI power by a further 5MW. The primary focus of the upgrade has therefore been to a) extend the pulse length of the two existing 2.5MW beamlines to 5s and b) modify one of the beamlines to deliver off-axis injection to the plasma."

Objective a) requires new beamline components, in particular new bend magnetics Residual Ion Dump (RID). The designs for these component have been completed and contracts placed for the key components. In addition, on site assembly of some of the key structural elements has begun (see figure 12). All mechanical hardware for the two MAST-U beamlines is expected to be completed by early 2015, well in time for integrating commissioning of the new machine.

Objective b) requires the raising of one of the beamlines 650mm above the mid-plane along with modifications to all its services (e.g. cooling, vacuum, gas, C&I etc). A key milestone was to remove and refurbish the NBI vessels and install a new support structure by May 2014. This was achieved ahead of schedule (see figure 13). As with the beamlines themselves, the new services are expected to be completed in time for integrated commissioning.

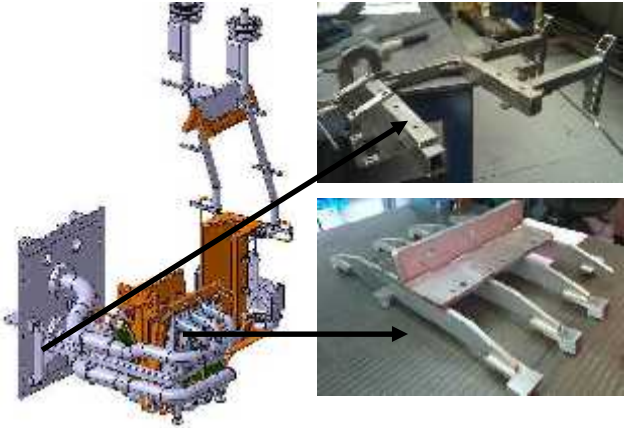

Fig. 12. MAST-U NBI Residual Ion Dump (left) with progress made in support structure manufacture (right)

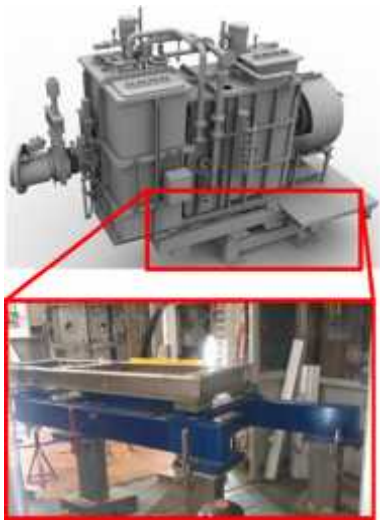

Fig. 13. Neutral Beam Support Structure required to reconfigure for off-axis injection

## 2.6 Machine Control

The complexity of the new machine necessitates an overhaul of the machine control system, including the central control system, machine protection, plasma control and data acquisition systems. The plasma control system in particular has to treat a larger number of analogue signals (more divertor coils and more gas valves). Methods of communicating this signals digitally over fibre have been studied and new systems using PCIe remote extenders and also bespoke Field Programmable Gate Array (FPGA) based analogue interface boards using low latency 1GbE UDP links have been developed. A single UDP packet can be distributed by multicast to numerous remote actuators via a single switch on a private network to achieve low latency control.

The vertical plasma control must be done with very low latency and this is also performed using an adaptive algorithm to accommodate ELMs using an FPGA based processing unit. The FPGA provides over optical fibre the class D signals required for the new radial field amplifier.

The larger magnetic fields produced by the new coils lead to sufficiently high load cases that significant mechanical damage could result from certain coil current combinations. A real time protection system is being developed to allow the processing of signals from around 100 sensors on remote FPGA units to be combined and analysed in a central FPGA based processing unit. The open hardware White Rabbit switch [9] developed at CERN has been chosen to perform this function. This device was designed to provide accurate timing synchronisation between distributed units via 18 SFP ports of GbE, however in this application the FPGA is completely re-programmed to communicate via the Aurora protocol of 1.25 Gbps fibre optic links and to perform the protection calculations necessary to generate the array of shutdown signals relevant to a catalogue of potential limiting scenarios. The use of programmable FPGA hardware leads to a high performance and cost effective solution.

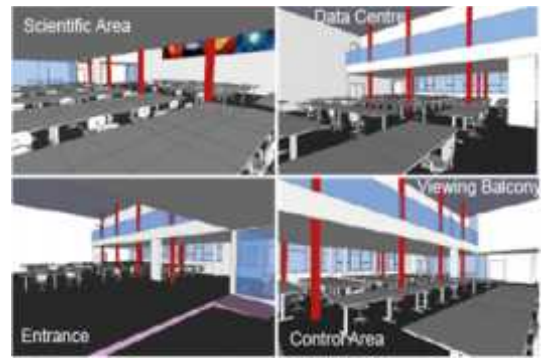

Fig. 14. Architect impression of upgraded MAST-U Control Room showing control, scientific and entrance areas

Following a detailed review in 2013 of the performance measures of the old MAST control room, the project decided to build a new two-storey control room and data centre for the new machine (see figure 14). The design includes many enhancements such as the marked increase in the size of functional areas, improved group display systems, a separate breakout room, a first floor viewing balcony and raised access flooring with flexible data and power infrastructure throughout so as to allow a flexibility in the layout of workstations. It is expected this new control room will be fully operational by mid 2015.

### 3. Lessons Learned thus far

Whilst MAST-U is a Medium-Sized Tokamak and lacks some of the complexity of other devices recently constructed or under construction, this is still a once-in-a-decade opportunity for CCFE to develop a team capable of building a fusion device allowing them to make more valuable contributions to other tokamak design / build programmes around the world (ITER, JT-60A, DEMO etc). There are also a number of lessons learned which should be fed back to the community to enhance their chances succeeding in their own programmes.

Most of the lessons learned thus far on this project can broadly be grouped into two areas: Project Management and Systems Engineering. These are summarized below.

### 3.1 Project Management Lessons Learned

- Ensure the project team has sufficient support in the area of cost estimating
- Hire in people from industry to work in design / procurement team (this brings in manufacturing experience, access to networks, fresh look at fusion engineering challenges)
- Ensure Value Engineering considered as early as possible in design process
- Implement tools to properly assess and report design maturity (e.g. detailed CAD model does not equal Design Complete)
- Co-locate the design team where-ever possible; this helps to make up for shortfalls in interface management system (no system is perfect)
- Avoid over-centralising approvals; seek to empower key staff and delegate technical / managerial / budgetary responsibilities down the management chain as much as possible
- Long timescales in fusion result in a lack of continuity in engineering delivery and experience. Inexperienced teams repeat mistakes if the knowledge isn't captured. Capturing lessons learned and knowledge gained from this and other fusion builds would be highly beneficial.

### 3.2 Systems Engineering Lessons Learned

- Allow time and resources up front to ensure Requirements cascade from top level User Requirements right through to System and Sub-system Requirements.
- Put in place a robust configuration control process early in the project, including a well-defined Plant Breakdown Structure and common terminology (this can be challenging in a project such as this where legacy plant is involved)
- Define geometric dimensioning and tolerancing early in the design process
- Implement efficient processes to quickly agree sensible compromises between maximum performance versus simplicity / reliability / cost.
- Rely on more smaller scale prototypes

### 4. Summary and Next Steps

At the time of writing, construction of MAST-U was projected to be completed mid-2015, followed by a period of integrated commissioning to deliver first plasma in early 2016.

In parallel with the build and commissioning activities, CCFE continues to actively seek out new collaborators to come and exploit this brand new facility. With detailed planning of the first experimental campaigns underway, it is clear that this new machine will be in a position to make vital contributions to the global fusion programme throughout the $2^{nd}$ half of this decade and beyond.

### Acknowledgments

This work was funded by the RCUK Energy Programme under grant EP/I501045